\documentclass[a4paper,doublecol,figures]{epl2}

\usepackage{graphicx}
\usepackage{dcolumn}
\usepackage{bm}
\usepackage{dsfont}
\usepackage{color}

\usepackage{url}
\usepackage[colorlinks=true ,citecolor=blue,urlcolor=blue]{hyperref}
\usepackage{amsmath,amssymb}

\def\txt#1{\textnormal{#1}}

\newcommand{\eqn}{Eq.~}

\newcommand{\ii}{\text{i}}
\newcommand{\ee}{\text{e}}
\newcommand{\vv}[1]{\boldsymbol{#1}} 

\newcommand    {\spinup}     {\mathord{\uparrow}}
\newcommand    {\spindown}   {\mathord{\downarrow}}

\begin{document}

\title{Topological phase transitions between chiral and helical spin textures in a lattice with spin-orbit coupling and a magnetic field}
\shorttitle{Topological phase transitions between chiral and helical spin textures}

\author{N.~Goldman\inst{1}\thanks{E-mail: \email{ngoldmanATulb.ac.be}} \and W.~Beugeling\inst{2} \and C.~Morais Smith\inst{2}}
\shortauthor{N.~Goldman \etal}
\institute{
\inst{1} Center for Nonlinear Phenomena and Complex Systems - Universit\'e Libre de Bruxelles (U.L.B.), B-1050 Brussels, Belgium\\
\inst{2} Institute for Theoretical Physics, Utrecht University - Leuvenlaan 4, 3584 CE Utrecht, The Netherlands
}


\abstract{We consider the combined effects of large spin-orbit couplings and a perpendicular magnetic field in a 2D honeycomb fermionic lattice. This system provides an elegant setup to generate versatile spin textures propagating along the edge of a sample.  The spin-orbit coupling is shown to induce topological phase transitions between a helical quantum spin Hall phase and a chiral spin-imbalanced quantum Hall state. Besides, we find that the spin orientation of a single topological edge state can be tuned by a Rashba spin-orbit coupling, opening an interesting route towards quantum spin manipulation. We discuss the possible realization of our results using cold atoms trapped in optical lattices, where large synthetic magnetic fields and spin-orbit couplings can be engineered and finely tuned. In particular, this system would lead to the observation of a time-reversal-symmetry-broken quantum spin Hall phase.}

\maketitle

\section{Introduction}

\begin{figure}[h!]
	\centering
	\includegraphics[width=1\columnwidth]{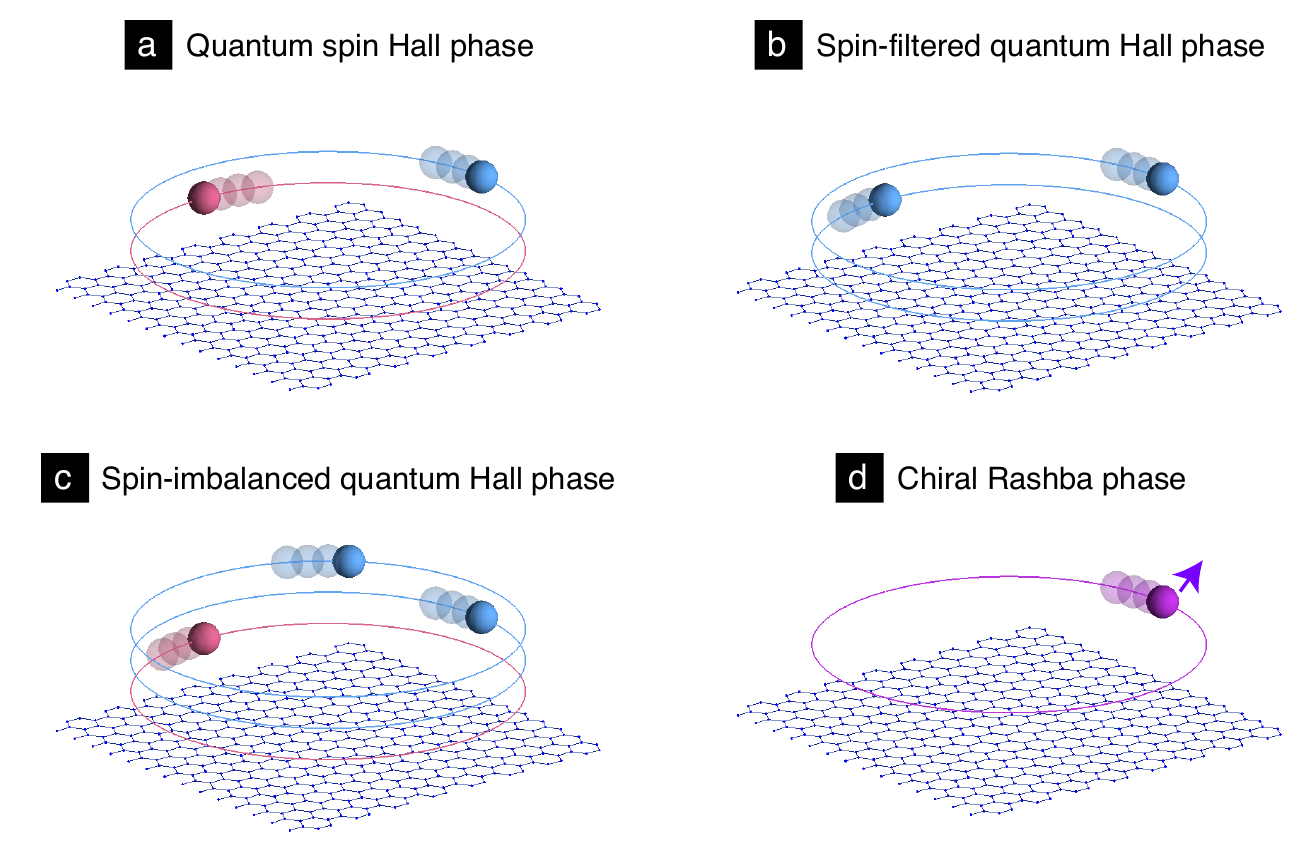}
	\caption{\label{fig1} {\bf Spin-textures and quantum Hall phases:}
	Spin-up and spin-down components are represented by blue and red balls, respectively. {\bf (a)} The QSH phase: the helical edge states are formed by counter-propagating spins of opposite sign. {\bf (b)} The spin-filtered quantum Hall phase: the chiral edge states are constituted of identical spins with same orientation. {\bf (c)} The spin-imbalanced QH phase: the chiral edge states feature several spins of different nature such that $N_{\uparrow} \ne N_{\downarrow}$. {\bf (d)} The chiral Rashba phase: a single chiral edge state is formed with arbitrary spin direction, represented in purple.}
\end{figure}

The discovery of the first topological state of matter, the integer quantum Hall (QH) effect~\cite{VonKlitzing1986}, has revolutionized our understanding of condensed matter: a two-dimensional (2D) electron gas subjected to a perpendicular magnetic field is gapped and insulating in the bulk, but features chiral edge states that carry quantized charge currents. The quantization of the Hall conductance $\sigma_H= \nu (e^2/h)$ is directly related to the number of these chiral edge states $\nu$, which is a sum of topological Chern numbers \cite{Kohmoto1985}. The proposal by Kane and Mele
\cite{KaneMele2005PRL95-14}, that a new topological state of matter could be realized in graphene by taking into account spin-orbit interactions, has brought about the quantum spin Hall (QSH) effect, in which the edge currents carry spin but no charge. This spin transport is realized  by helical edge states, namely spins traveling in opposite directions along the edge [cf.\ Fig.~\ref{fig1} (a)], and is protected by a $\mathbb{Z}_2$ topological index \cite{KaneMele2005PRL95-14}.  Contrarily to the integer QH effect, which is rooted in a time-reversal symmetry (TRS) breaking perturbation (i.e., a magnetic field), the QSH effect is a consequence of the so-called intrinsic spin-orbit coupling (ISO). The latter acts as opposite ``fluxes" \cite{Haldane1988} on the two spin components. Thus, the system effectively exhibits two QH phases, which get transformed into each other by time reversal, giving rise to helical edge states.

The two topological states of matter described above produce either charge or spin edge currents that are dissipation-free and robust against disorder, due to their topological character. In conventional semiconductors, which exhibit the integer QH effect, the spin texture of the edge states is fixed by the competition between the cyclotron energy and the Zeeman spin splitting produced by the external magnetic field. In particular, the observation of Hall plateaus at high odd fillings $\nu$ requires spin-resolved Landau levels which can be achieved by enhancing the Zeeman energy, e.g., by tilting the sample \cite{HwangEA1993} or by engineering materials with large Land\'e $g$ factors \cite{ButtnerEA2011}. In the limit of extremely large Zeeman energy, one eventually expects to observe spin-filtered QH states, where all the current-carrying edge states have the same spin orientation [cf.\ Fig.~\ref{fig1} (b)].

An interesting question is whether topological phase transitions between helical QSH and chiral QH phases, exhibiting distinct spin textures on the edge, could be driven in a system combining the effects of both a magnetic field and spin-orbit couplings. In this Letter, we address this fundamental problem by solving the Kane-Mele model \cite{KaneMele2005PRL95-14} in the presence of an external magnetic field. Although it is an interesting model from a theoretical point of view, its physical realization as a model for graphene has strong limitations: the topological phases emanating from the Kane-Mele model cannot be observed experimentally with graphene because of its extremely small ISO coupling \cite{BoettgerTrickey2007}. The physics of this model can however be studied within an alternative and promising framework offered by cold atoms trapped in optical lattices. Indeed, controllable Abelian and non-Abelian synthetic gauge fields can mimic the effects of magnetic fields and SO couplings in a highly versatile experimental framework \cite{LinEA2009,LinEA2011Nature,DalibardEA2010preprint,GoldmanEA2010,LimEA2010,AlbaEA2011preprint}. In addition, a Zeeman splitting term can also
be easily engineered \cite{MakogonEA2010preprint}.
These recently developed technologies offer the unique possibility to explore the regimes of extremely large magnetic field and spin-orbit couplings. Moreover, cold-atom systems allow us to consider quantum phases in the absence of disorder, in which case the QSH state should remain stable even in the presence of TRS breaking perturbations.

Until now, most of the proposals for realizing QH lattice models with cold atoms are based on the spinless Haldane or Hofstadter models \cite{JakschZoller2003,StanescuEA2010}. In these configurations, one expects to generate spin-filtered QH states, since only one spin component (i.e., internal atomic state) is considered in the dynamics. On the other hand, an $\mathrm{SU}(2)$ generalization of the Hofstadter model, using more internal states, has been introduced in Ref.~\cite{GoldmanEA2010} to engineer the QSH phase. However, despite their ability to either exhibit the QH or QSH phases, the lattice models of Refs. \cite{JakschZoller2003,StanescuEA2010,GoldmanEA2010} do not fully capture the interplay between quantum Hall physics and the lifting of the spin degeneracy, present in condensed-matter systems due to Zeeman and spin-orbit effects.

In this Letter, we show that a two-component lattice combining Zeeman and spin-orbit terms constitutes an ideal playground to produce versatile topological insulating states with rich spin textures. It is shown that an optical-lattice realization of this system could display remarkable phase transitions between chiral and helical spin structures, not envisaged so far. The synthetic magnetic field and spin-orbit terms are generated by inducing a gauge field in the atomic cloud that influences all the hopping terms \cite{LinEA2009,JakschZoller2003,DalibardEA2010preprint,GoldmanEA2010,LimEA2010}, whereas the Zeeman splitting filters the spins. Different combinations of these terms will provide us with novel topological phase transitions that modify the spin texture characterizing the edge states. Interestingly, these transitions are driven by the spin-orbit couplings, while the Fermi energy and external magnetic field are held fixed. Furthermore, we show that this setup allows to manipulate the spin orientation of an isolated edge state, thus generating richer spin structures than in standard topological insulating phases.

\section{Results}

We consider the Kane-Mele model, which describes spin-$1/2$ fermions in a honeycomb lattice, subjected to a uniform perpendicular magnetic field $ \vv{B}=B \vv{1}_z$. The presence of the magnetic field modifies the graphene model of Ref.~\cite{KaneMele2005PRL95-14} in two fundamentally different manners. First, the magnetic field acts on the charges through the minimal coupling $\vv{p}-e \vv{A}$, where $\vv{A}$ is the gauge potential, and enters the model through the so-called Peierls phases \cite{Hofstadter1976}. Then, an additional coupling leads to a constant Zeeman splitting $g \mu_B B \hat{\sigma}_z$, where $g$ is the Land\'e factor, $\mu_B$ is the Bohr magneton  and $\hat{\sigma}_z$ is the Pauli matrix acting on the spins. The tight-binding Hamiltonian of interest for our work then takes the following form
\begin{align}\label{eq:ham}
  &\mathcal{H}
  = - t \sum_{\langle k,l \rangle} \ee^{\ii \phi_{lk}}c_{k}^\dag  c_{l}
    + \ii   \lambda_\text{SO} \sum_{\langle\!\langle k,l \rangle\!\rangle}
      \nu_{k,l} \ee^{\ii \phi_{lk}} c_{k}^\dag \hat{\sigma}_z c_{l}  \notag\\
  & + \ii   \lambda_\text{R} \sum_{\langle k,l \rangle} \ee^{\ii \phi_{lk}} c_{k}^\dag (\hat{\vv{\sigma}} \times \mathbf{d}_{kl})_z  c_{l}
    + 2 \pi \Phi \lambda_\text{Z} \sum_{k} c_{k}^\dag \hat{\sigma}_z c_{k}.
\end{align}%
\begin{widetext}[!]
\begin{align}
&E  \Psi_n
  = \mathcal{D}_n  \Psi _n\!+\!  \mathcal{R}_n \Psi_{n+1}\! +\! \mathcal{R}^{\dagger}_{n-1} \Psi_{n-1} , \notag \\
& \mathcal{D}_n
=\begin{pmatrix}
2  \lambda_\text{SO} \hat{\sigma}_z \sin \bigl ( 2 \pi \Phi(n+ \frac{1}{6})+k\bigr )+2 \pi \Phi  \lambda_\text{Z}\hat{\sigma}_z&t \hat{1}- \ii \lambda_{\text{R}} \hat{\sigma}_y \\
t \hat{1} + \ii \lambda_{\text{R}} \hat{\sigma}_y &-2  \lambda_\text{SO} \hat{\sigma}_z \sin \bigl ( 2 \pi \Phi(n+ \frac{1}{6})+k\bigr ) +2 \pi \Phi  \lambda_\text{Z}\hat{\sigma}_z\\
\end{pmatrix}, \notag \\
& \mathcal{R}_n
=\begin{pmatrix}
\ii  \lambda_\text{SO} \hat{\sigma}_z  \bigl (\ee^{\ii\pi\Phi(n+\frac{2}{3})}-\ee^{-\ii\pi\Phi(n+\frac{2}{3})-\ii k} \bigr )& 0 \\
\ee^{\ii\pi\Phi(n+1)} \bigl (t \hat{1}- \ii \lambda_{\text{R}} \hat{\gamma}_{-} \bigr ) + \ee^{-\ii\pi\Phi(n+1)-\ii k} \bigl( t \hat{1}- \ii \lambda_{\text{R}} \hat{\gamma}_{+}  \bigr ) & - \ii  \lambda_\text{SO} \hat{\sigma}_z  \bigl (\ee^{\ii2\pi\Phi(n+\frac{7}{6})}- \ee^{-\ii2\pi\Phi(n+\frac{7}{6})-\ii k} \bigr )\\
\end{pmatrix},\notag\\
&\hat{\gamma}_{\pm}= \pm \frac{\sqrt{3}}{2} \hat{\sigma}_x + \frac{1}{2} \hat{\sigma}_y.
\label{eq:single}
\end{align}
\end{widetext}%
Here, $c_{l}^{(\dagger)}=(c_{l,\spinup},c_{l,\spindown})^{(\dagger)}$ are the annihilation (creation) operators on the lattice site $l$ for fermions with spin components up and down. The first two terms  in \eqn\eqref{eq:ham} correspond, respectively to nearest (NN) and next-nearest (NNN) neighbour hopping on the honeycomb lattice, with the associated amplitudes $t$ and $\lambda_\text{SO}$. The second term is due to the ISO coupling and is characterized by spin-dependent amplitudes, with  $\nu_{k l}=\pm 1$ according to the  orientation of the path connecting the NNN sites $k$ and $l$. Here, we set $\hbar=e=1$ and we use the hopping parameter $t$ and the bond length $a\equiv1$ as the energy and length units, respectively. The Peierls phases, $\phi_{lk}=\ii \int_l^k \vv{A} \cdot \vv{\text{{\bf d}} l}$, act on both the NN and NNN hopping terms and are generally expressed in terms of the dimensionless parameter $\Phi=S_6 B /2\pi$, i.e., the number of magnetic flux quanta per plaquette, where $S_6= 3 \sqrt{3}/2$ is the area of a plaquette.  The third term corresponds to a Rashba SO coupling, which acts as a spin-mixing perturbation between NN sites: it tends to misalign the spin by rotating it towards the plane of the sample ~\cite{KaneMele2005PRL95-14}. In this term, $\lambda_\text{R}$ is the coupling strength, $\hat{\vv{\sigma}}$ is the vector of Pauli matrices and $\mathbf{d}_{kl}$ is the vector between sites $k$ and $l$. The last term in \eqn\eqref{eq:ham} is the constant Zeeman splitting, where we set  $\lambda_\text{Z}=g \mu_B S_6^{-1}$. The potential realization of the model \eqref{eq:ham} with ultracold atoms trapped in optical lattices allows us to envisage the case of extremely large magnetic fluxes $\Phi \sim 0.1-1$, where the spectral gaps $\Delta \sim 0.1-1t$ lead to Hofstadter fractal structures \cite{Hofstadter1976,Rammal1985}. In contrast to condensed-matter experiments (in which $\lambda_\text{Z}$ is basically fixed by the Land\'e factor), optical lattice setups allow us to tune every parameter individually to arbitrary values. In this framework,  exotic configurations could be realized, e.g., featuring large magnetic flux and spin-orbit couplings while keeping the Zeeman splitting low. Furthermore, the non-interacting limit can be easily reached through Feshbach resonances \cite{BlochEA2008}.

In the following, we discuss novel topological phase transitions and the manipulation of isolated edge states, which are obtained by tuning the Zeeman splitting and the SO terms. Although our calculations are performed for the honeycomb lattice, our predictions do not rely on this specific geometry: Any system exhibiting a Dirac-like dispersion relation, large Zeeman splitting and spin-orbit interactions will exhibit a similar behaviour.

The spectrum and  topological phases stemming from Hamiltonian \eqref{eq:ham} have been widely studied for two limiting cases.  In the absence of magnetic flux, $\Phi=0$, a finite ISO coupling opens a QSH gap at half-filling, with magnitude $\Delta=6 \sqrt{3} \lambda_\text{SO}$\cite{KaneMele2005PRL95-14}. In the spinless limit $\lambda_\text{Z}=\lambda_\text{SO}=\lambda_\text{R}=0$, a finite magnetic flux $\Phi$ leads to a multi-gap QH system \cite{HatsugaiEA2006}. In this Letter, we explore the \emph{combined} effects of spin-orbit couplings and magnetic flux on these QH and QSH gaps. Since the honeycomb lattice has two sites per unit cell, $A$ and $B$, the single-particle wave function can be written as $\Psi_{n,m}=\big ( \psi^A_{n,m}, \psi^B_{n,m} \bigr )$, where $(n,m)$ are integers labeling the unit cells \cite{HouYang2009}, and where the spin components are implicit. Using the Landau gauge, we set $\Psi_{n,m}=\exp(\ii k m) \Psi_n$ and the single-particle Schr\"odinger equation takes the form of a generalized Harper equation, \eqn\eqref{eq:single}.

Topological phases are distinguished by the quantum currents carried by gapless edge states, which are obtained by computing the spectrum $E=E(k)$ on an abstract cylindrical geometry \cite{Hatsugai1993}. The transport coefficients are equally evaluated through the computation of topological invariants, the Chern numbers and $\mathbb{Z}_2$ index, which are associated with the bulk energy bands \cite{Kohmoto1985,KaneMele2005PRL95-14}.

Let us first discuss the fate of the QSH phase as a uniform magnetic field is progressively turned on. The main effect of this external field is to break TRS and therefore to allow scattering processes between the counter-propagating edge states characterizing the QSH phase \cite{KaneMele2005PRL95-14}. Therefore, although the bulk energy gap and the non-trivial topological index associated with the QSH phase at half-filling are preserved for a finite magnetic flux, the TRS violation destroys the topological phase in the presence of disorder. Such phases, referred to as ``weak QSH" (WQSH) phases or TRS-broken QSH phases, are thus trivial in real materials, but they would still be robust in optical lattices where disorder is absent. We note that the robustness of the WQSH has recently been investigated in Ref.~\cite{YangEA2011}, for a system featuring a constant exchange term. Naturally, the magnetic field also creates additional spin-degenerate QH phases away from half-filling. For sufficiently large magnetic fields, the WQSH gap at half-filling is destroyed (e.g., when $\lambda_\text{SO}=0.05 t$ and $\lambda_\text{Z}=0.1t$, this gap closing occurs at $\Phi \approx 0.14$) and only the standard QH gaps remain. We highlight that the WQSH phase described here relies on the existence of a bulk energy gap,  characterized by a non-trivial $\mathbb{Z}_2$ index, which guarantees the robustness of this phase for finite magnetic flux (in the absence of disorder).

\begin{figure*}[t!]
	\centering
	\includegraphics[width=1.7\columnwidth]{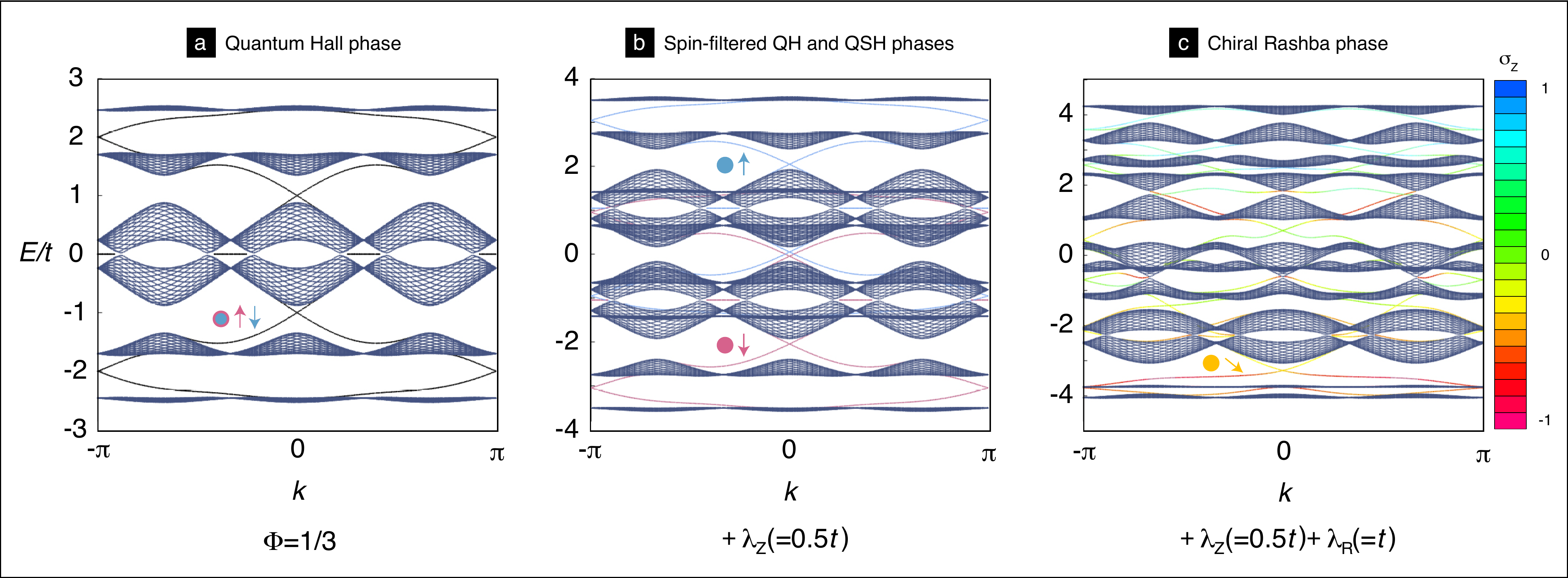}
	\caption{\label{figrashba} {\bf Generating a chiral Rashba phase on the edges:} The energy spectrum $E=E(k)$ is represented as a function of the momentum, in a cylindrical geometry with zigzag edges. {\bf (a)} Several QH phases are generated by a magnetic flux $\Phi=1/3$. When the Fermi energy lies in the gap at $E \approx - t$, a single spin-degenerate edge state propagates on each edge. {\bf (b)}  The inclusion of  the Zeeman splitting lifts the spin degeneracy and creates spin-filtered QH phases at $E=\pm 2.25 t$ and $E=\pm 3 t$. At half-filling, the Zeeman splitting opens a ``weak" QSH gap. {\bf (c)}  The addition of the Rashba term then allows for flipping the spin of the edge state to an arbitrary orientation. When the Fermi energy $E_{\text{F}} \approx -3.25 t$, the single edge state has spin components $\sigma_{z} \approx -0.5$ and $\sigma_{\bot} \approx 0.5$ (cf. orange dot). Here, the bulk energy bands are depicted in dark blue. }
\end{figure*}

\begin{figure*}[t!]
	\centering
	\includegraphics[width=1.7\columnwidth]{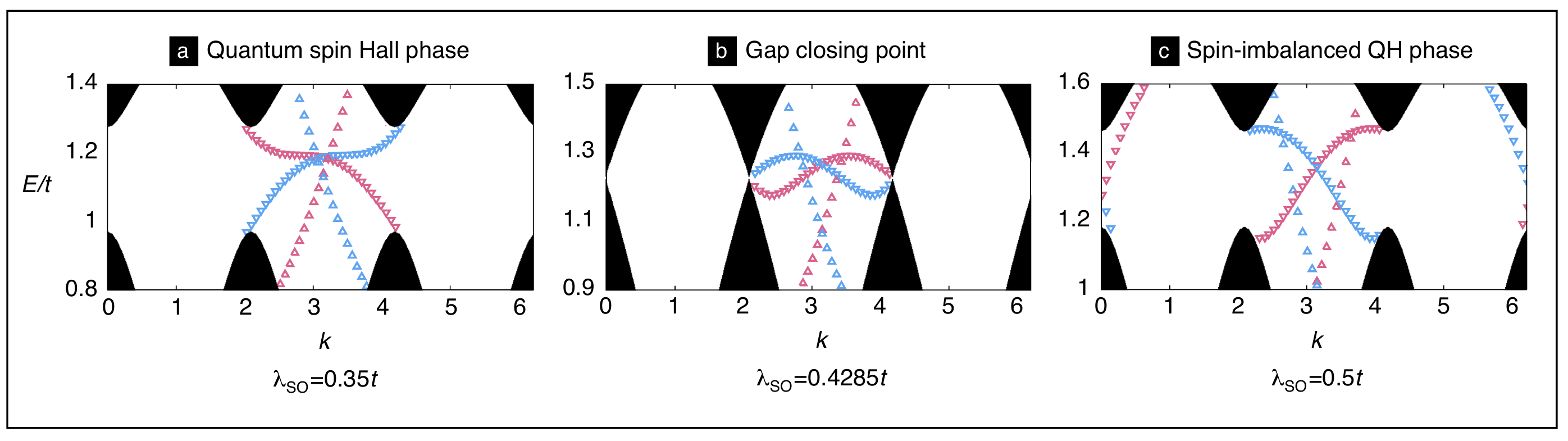}
	\caption{\label{figimb}  {\bf Phase transition from the QSH to the spin-imbalanced QH phase:} The energy spectrum $E=E(k)$ is represented as a function of the momentum, in a cylindrical geometry with bearded edges. Here we set $\Phi=1/3$ and $\lambda_\text{Z}=0.5t$. Spin up [resp. down] states are represented by up [resp. down] triangles and the two colours designate the two edges of the cylinder where the states are localized. {\bf (a)} The weak QSH phase is characterized by counter-propagating spins with opposite sign. {\bf (b)} The gap closes for $\lambda_\text{SO} \approx 0.4285 t$. {\bf (c)} The spin-imbalanced QH phase: each edge features one spin-up and two spin-down states propagating in the \emph{same} direction. Here, the bulk bands are depicted in black. }
\end{figure*}

We now investigate an alternative and more interesting scenario, starting from spin-degenerate QH phases in the absence of Zeeman splitting and spin-orbit coupling. The energy spectrum for $\Phi=1/3$ shows four spin-degenerate QH phases, with the Hall conductivities $\sigma_H=\{ 2, -2, 2, -2 \}$ in units of the conductivity quantum. Now, let us  turn on the Zeeman splitting: by lifting the spin-degeneracy, the Zeeman term separates the QH phases and eventually generates \emph{spin-filtered} QH phases [cf.\ Fig.~\ref{fig1} (b) and Fig.~\ref{figrashba} (b)]. When  the Fermi energy is located inside such a spin-filtered QH bulk gap, a two-component Fermi gas will exhibit edge currents transported by a \emph{single} species only.  Besides, the Zeeman term creates a WQSH phase at half-filling. It is remarkable that such a helical topological phase is produced in the absence of ISO coupling, which highlights the great similarity between the Zeeman and the ISO terms in \eqn\eqref{eq:ham} \cite{AbaninEA2011}. However, this Zeeman-induced QSH phase is weak because of the TRS violation [cf. discussion above]. Starting from this configuration, we progressively add the ISO coupling. As shown in Fig.~\ref{figimb} (a), the ISO opens new gaps in the energy spectrum,  the topological character of which highly depends on the Zeeman splitting. For $\lambda_\text{Z}=0.5t$ and $\lambda_\text{SO}=0.35t$, a WQSH gap  opens at $E \approx 1.1 t$ [cf.\ Fig.~\ref{figimb} (a)]. Increasing  the ISO term further leads to the closure of the gap, depicted in Fig.~\ref{figimb} (b), and eventually to a topological phase transition. As represented in Fig.~\ref{figimb} (c), the gap is now characterized by a strong \emph{spin-imbalanced} QH phase [cf.\ Fig.~\ref{fig1} (c)]. Indeed, the edge is populated by chiral edge states with different spin components: $N_{\uparrow}=1 \ne N_{\downarrow}=2$. In this configuration, when the Fermi energy $E_{\txt{F}}$ lies in the gap at $E \approx 1.3 t$, the Hall conductivity and the spin-Hall conductivity differ, since $\sigma_H=3$ and $\sigma_H^s=-1$ in units of the charge and spin conductivity quanta, respectively. Note that the chiral character of these current-carrying edge states guarantees the robustness of this quantum phase against external perturbations. In this sense, the ISO coupling has driven a transition from a \emph{weak} QSH to a \emph{strong} spin-imbalanced topological phase, while the Fermi energy and magnetic field were kept constant. Consequently, by adjusting the Zeeman splitting and the ISO coupling, one is able to tune the spin-imbalance $N_{\uparrow} \ne N_{\downarrow}$ characterizing the current-carrying edge states. Recently, one-dimensional spin-imbalanced atomic gases have been under investigations \cite{NascimbeneEA2009}, and lead to the realization of exotic pairing mechanisms. Our 2D model features spin-imbalanced states that are confined to the 1D edge of the system, but also possess a chiral nature. We note that  spin-imbalanced QH phases have already been observed in solid-state laboratories: In semiconductor heterostructures without ISO coupling (e.g., \chem{GaAs/AlGaAs}), a transition from $\nu = 2$ (i.e., $N_{\uparrow}= N_{\downarrow}=1$) to the spin-imbalanced phase $\nu = 3$ (i.e., $N_{\uparrow}= 2$ and $N_{\downarrow}=1$) is driven by tuning the magnetic field. On the contrary, in our model the ISO interaction drives the transition, while the magnetic field is held constant. Furthermore, the variation of the magnetic field in semiconductors causes a phase transition between two chiral states, whereas in our study the ISO drives a transition between the WQSH phase, which is helical, to the chiral spin-imbalanced QH phase. Thus, the interplay between spin-imbalance, chirality and controllable interactions should lead to rich physics not envisaged so far.

The Rashba SO coupling has dramatic consequences on the system described by \eqn\eqref{eq:ham}, as it induces spin mixing. In the presence of a magnetic field, the Rashba term lifts the spin degeneracy of the original QH phases and eventually opens new topologically \emph{trivial} gaps. The Rashba term therefore \emph{destroys} the topological nature of the QH gaps in certain regions of the spectrum (cf.\ Ref.\ \cite{KaneMele2005PRL95-14} without magnetic field).  In addition, the Rashba coupling partially polarizes the edge currents associated with the QH phases.

A remarkable spin-manipulation process can be envisaged by combining the effects of the Rashba spin-mixing perturbation and the Zeeman splitting. As already mentioned above, the Zeeman splitting generates spin-filtered edge states in the QH gaps. In the example illustrated in Fig.~\ref{figrashba} (b), a single spin-down particle propagates along the edge of the sample when $E_{\txt{F}} \approx -2.25 t$, for $\Phi=1/3$ and $\lambda_\text{Z}=0.5 t$. As shown in Fig.~\ref{figrashba} (c), one can then progressively increase the Rashba coupling and control the spin orientation $\langle\sigma_{\mu}\rangle=   \langle \Psi \vert  \hat{\sigma}_{\mu} \vert \Psi \rangle$ of this specific and \emph{unique} chiral edge state. This process indeed modifies the band structure, as well as the spin textures associated  with the edge current, leading to the chiral Rashba phase illustrated in Fig.~\ref{fig1} (d): the spin components of this single edge state are now $\sigma_{z}=-0.5$ and $\sigma_{\bot}=0.5$, where $\sigma_{\bot}$ is the spin component in the direction perpendicular to both the direction of propagation and the applied magnetic field. The canting of the edge-state spin is a hallmark of the competition between the Rashba term, which tends to rotate the spin direction towards the $xy$ plane, and the Zeeman term, which tends to align the spins in the vertical direction. For $\lambda_\text{R}=t$, a large but realistic value for a Raman-induced tunneling amplitude, this single edge state can be reached by adjusting the Fermi energy to the value $E_{\text{F}} \approx -3.25 t$, as shown in Fig.~\ref{figrashba} (c).  Interestingly, the spin direction of the edge states within this gap highly depends on the Fermi energy \cite{Rashba2009}. Thus, the spin direction of the edge states can be tuned by varying the filling factor. We emphasize that this effect is notably different from the constant spin orientation that would result from the Zeeman effect in a tilted magnetic field.  We finally stress that the variation of $\lambda_\text{R}$ does not close the topological bulk gap, and therefore, its associated Chern number remains unaltered. In this sense, there is no topological phase transition involved in this manipulation of the edge-state spin.

\section{Discussion and conclusions}

The experimental realization of the topological phase transitions presented here constitutes an appealing target for cold-atom laboratories. In principle, one could realize the model described by \eqn\eqref{eq:ham} by trapping ultracold fermions in a honeycomb optical lattice. Indeed, this geometry has been recently realized experimentally, but we stress that simpler geometries with additional pseudo-spin degrees of freedom would lead to similar effects \cite{LimEA2010}. The main elements that have to be added to these lattice systems are the uniform magnetic field and spin-orbit couplings, which constitutes a difficult but appealing task. Both effects have already been engineered in cold gases confined to a single trap, by using Raman transitions between internal atomic states \cite{LinEA2009}. However, different methods are required for generating such fields in atomic {\it lattices}, as have already been suggested in Refs. \cite{JakschZoller2003,GoldmanEA2010,BermudezEA2010}. In particular, we note that such a method has been recently described for the honeycomb lattice \cite{GoreckaEA2011}. These proposals could be adjusted to generate the Rashba and ISO terms, as well as the Peierls phases. Moreover, the Zeeman splitting in \eqn\eqref{eq:ham} could be easily produced by static magnetic fields \cite{MakogonEA2010preprint}. In this elaborate but fruitful configuration,  the parameters $\lambda_\text{SO}$, $\lambda_\text{R}$, and $\lambda_\text{Z}$ are finely tuned by external fields. This high control over the parameters allows to drive the transitions proposed in this work.
A difficulty encountered in cold-atom realizations of topological insulating phases is the presence of confining traps, which will generally alter the bulk bands and destroy the edge-state structures \cite{StanescuEA2010}. To overcome this drawback, it is possible to either design sharp walls or to create interfaces in the system  \cite{GoldmanEA2010}. We also stress that the topological phases presented in Figs.~\ref{figrashba} and \ref{figimb} are protected by bulk energy gaps $\Delta \sim 0.1 - 1 t$, thus requiring achievable temperatures $T \sim 10\un{nK}$ for their observation.
It is worth mentioning that the possibility to tune the inter-particle interactions with Feshbach resonances, combined with controllable Zeeman splitting and synthetic Rashba SO coupling, would lead to a cold-atom realization of a topological superconductor exhibiting Majorana fermions \cite{SauEA2011}.

Finally, let us comment on the possibility to explore our results in a condensed-matter context, in which  achievable magnetic fields typically belong to the low-flux regime ($\Phi\ll1$). Recently, materials featuring strong ISO coupling and large Zeeman splitting, such as \chem{Hg(Cd)Te} quantum wells, have been developed in order to detect the QSH \cite{KonigEA2007,BruneEA2011} and spin-resolved QH states at low fields \cite{ButtnerEA2011}. Moreover, these setups allow experimentalists to study the competition between the Rashba SO and the Zeeman splitting \cite{GuiEA2004}. In principle, the effects discussed in this Letter, which result from the combination of spin-orbit couplings and magnetic fields, could be obtained in these arrangements for realistic values of the model parameters. For example, manipulating the spin orientation of spin-filtered QH edge states could be achieved by tuning the electric field responsible for the Rashba coupling and by varying the Fermi energy within the bulk gap (cf.\ Fig.~\ref{figrashba}).  However, these systems are unsuitable for detecting the topological phase transitions between helical WQSH and chiral QH phases. Indeed, a finite magnetic field breaks TRS and should therefore radically destabilize the WQSH phase in the presence of disorder: although the bulk gap and its non-trivial $\mathbb{Z}_2$ index might be preserved, the magnetic field triggers scattering processes between the counter-propagating (helical) edge states. Therefore, the WQSH phase induced by the Zeeman splitting (cf.\ Fig.~\ref{figrashba}) should reduce to a normal insulator in a solid-state experiment \cite{KaneMele2005PRL95-14}. Moreover, we note that the ISO interaction, which drives the WQSH-QH transition (cf. Fig. \ref{figimb}), cannot be varied in real materials. In conclusion, cold atoms subjected to synthetic gauge fields and Zeeman splitting will certainly provide a richer framework to study these interesting topological phase transitions and spin manipulation.

\acknowledgments

NG thanks the F.R.S-F.N.R.S for financial support. WB and CMS are supported by the Netherlands Organisation for Scientific Research (NWO). The authors acknowledge A. Bermudez, I. B. Spielman, F. Gerbier, J. Dalibard and M. Lewenstein for inspiring discussions.


\end{document}